\documentclass{ws-procs9x6}
\usepackage{epsfig}
\usepackage{amsmath}
\usepackage{wrapfig}
\begin{document}

\def\Journal#1#2#3#4{{#1} {\bf #2} (#3) #4}
\def\NCA{\em Nuovo Cimento}
\def\NIM{\em Nucl. Instrum. Methods}
\def\NIMA{{\em Nucl. Inst. Meth.} {\bf A}}
\def\NPB{{\em Nucl. Phys.}   {\bf B}}
\def\PLB{{\em Phys. Lett.}   {\bf B}}
\def\PRL{{\em Phys. Rev. Lett.}}
\def\PRD{{\em Phys. Rev.}    {\bf D}}
\def\ZPC{{\em Z. Phys.}      {\bf C}}
\def\EJC{{\em Eur. Phys. J.} {\bf C}}
\def\CPC{{\em Comp. Phys. Commun.}}

\title{Jets in Deep-Inelastic Scattering at HERA and determinations of \boldmath{$\alpha_s$}\footnote{\uppercase{T}alk given on behalf of the \uppercase{H}1 and
\uppercase{ZEUS} collaborations at the \uppercase{L}ake 
\uppercase{L}ouise \uppercase{W}inter \uppercase{I}nstitute 15-21~\uppercase{F}eb. 2004, 
\uppercase{L}ake \uppercase{L}ouise, \uppercase{C}anada}}

\author{R.~P\"OSCHL
}

\address{DESY Hamburg \\
Notkestr. 85, \\ 
22607 Hamburg, Germany\\ 
E-mail: poeschl@mail.desy.de}



\maketitle

\abstracts{
Several methods 
to extract the strong coupling constant $\alpha_s$ by means of highly energetic
jets in Deep-Inelastic Scattering are presented. The results from
the various methods agree with one another and with the world average. The errors are
competetive to those achieved in $\alpha_s$ determinations in other 
processes such as proton--anti-proton scattering. 
}

\section{Introduction}
Measurements of the hadronic final state in deeply inelastic $ep$
scattering (DIS) provide precision tests of 
quantum chromodynamics (QCD). At HERA data are collected over a 
wide range of the negative four-momentum-transfer $Q^2$, 
and the transverse energy $E_{T}$ of hadronic final state jets. 
As sketched in Eq.~\ref{Eq:jetcross} the jet cross section can be expressed
as a power series of $\alpha_s$ combined with a convolution of the
hard matrix element, $\hat{\sigma}_{jet}$, and appropriate parton distribution functions of the proton.
\begin{equation}
\sigma_{jet} = \sum\alpha^{n}_s(\mu_r)\sum\hat{\sigma}_{jet}(\mu_r,
    \mu_f)\otimes {\rm pdf}(\mu_f, ...),
\label{Eq:jetcross}
\end{equation}
with $\mu_r$ and $\mu_f$ being mass scales.

\begin{figure}[t]
\center
\epsfig{file=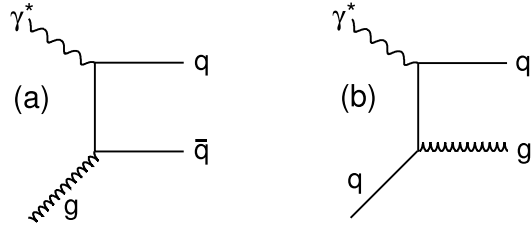,width=0.6\textwidth}
\caption{
Leading order diagrams for dijet production in $ep$ scattering. (a)
photon-gluon fusion and (b) QCD-Compton process. }
\label{fig:lo} 
\end{figure}

Fig.~\ref{fig:lo} shows diagrams of the leading order, here $O(\alpha_s)$, 
processes for dijet-production in DIS. 

The accurate measurement of jet
production, hence, allows for a precise determination of $\alpha_s$.

\section{Single Inclusive Jet Cross Section}  
\begin{figure}[bt]
\unitlength1cm
\begin{minipage}[t]{5.5cm}
\begin{picture}(6.5,6.5)
\put(0.3,0.8){\epsfig{file=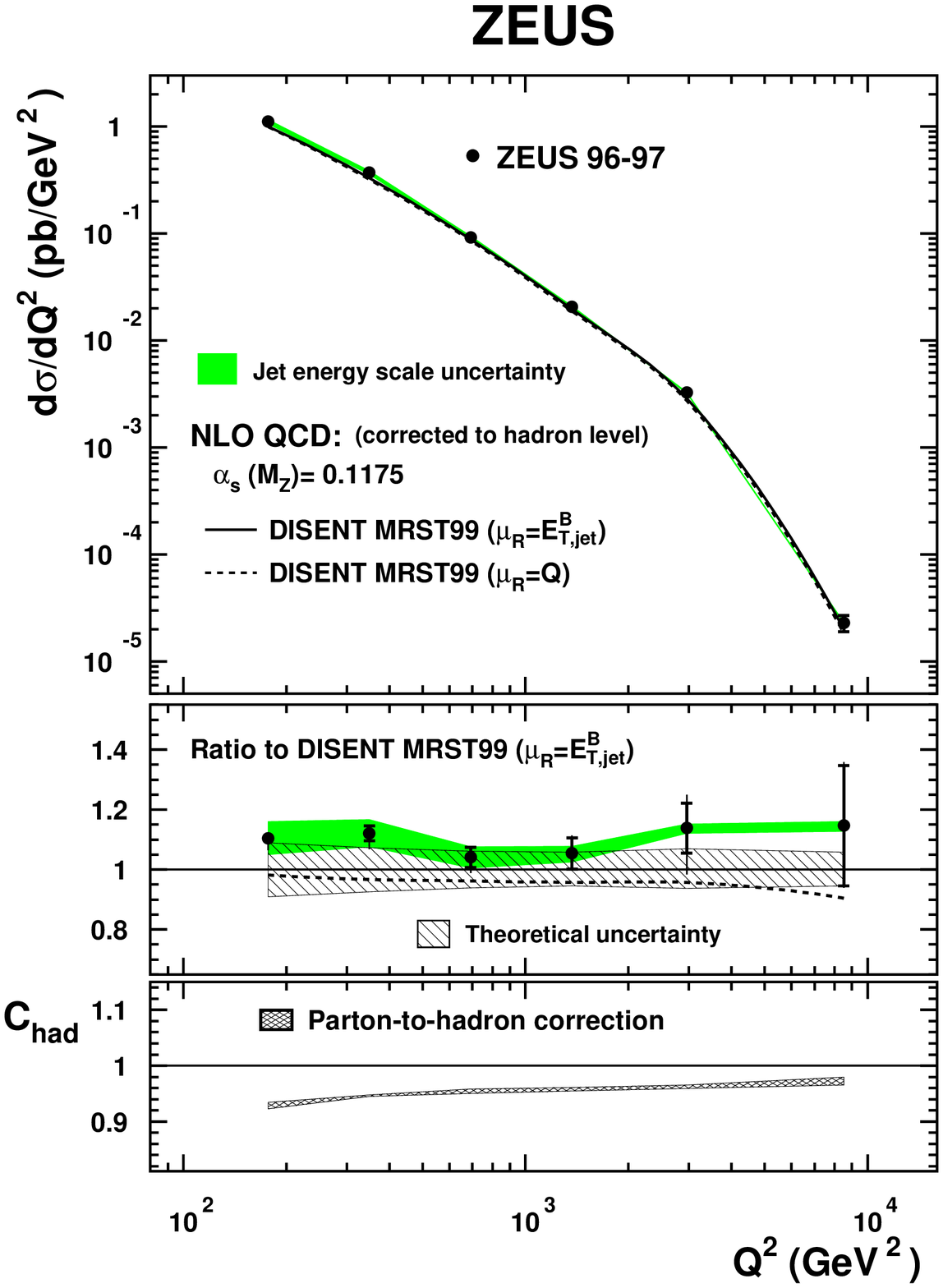,width=0.9\textwidth}}
\end{picture}\par
\end{minipage}
\hfill
\begin{minipage}[t]{5.5cm}
\begin{picture}(6.5,6.5)
\put(0.3,0.8){\epsfig{file=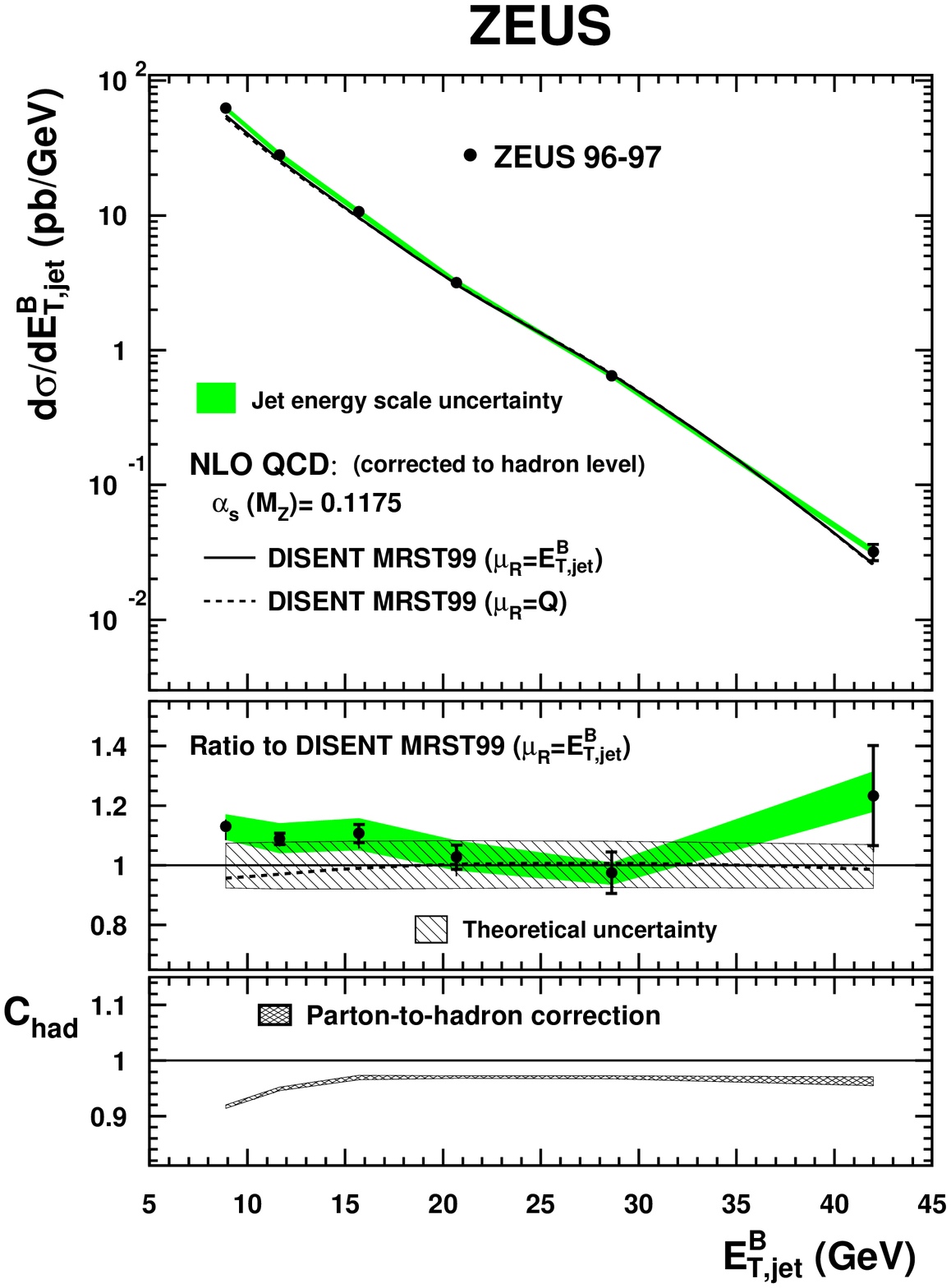,width=0.9\textwidth}}
\end{picture}\par
\end{minipage}
\vspace{-1.0cm}
\caption{
Single inclusive jet cross section as a function of
$Q^2$ (left) and $E^{B}_{T,{\rm jet}}$ (right) compared with NLO QCD 
predictions.} 
\label{fig:sicross} 
\end{figure}

For this analysis it is required to identify at least one jet above 
a given transverse energy.  Fig.~\ref{fig:sicross} shows the measured 
single inclusive jet cross section as a function of
$Q^2$ and the transverse energy of the jet as measured in the
Breit-Frame compared with NLO-QCD calculations~\cite{zeus:incl}. 

The data have a
typical experimental uncertainty of 7~\% and are well reproduced by
the theoretical predictions at large values of $E_T$ and $Q^2$. The NLO-QCD
calculations, currently only available at the 
parton level, are corrected for hadronisation effects using LO models.
These corrections are expected to be small at large $E_T$ and $Q^2$.
\subsection{Determination of $\alpha_s$} 
The dependency of a generic jet cross section on $\alpha_s(M_Z)$ is
parameterized in the corresponding analysis bins with the help of
suitable NLO-QCD predictions featuring
slightly different values of $\alpha_s(M_Z)$. By comparison of the 
parameterized cross section with the measured cross section the values 
for $\alpha_s$ are obtained. The resulting $\alpha_s(M_Z)$ from {\it e.g.} 
the single  inclusive jet cross section for $Q^2>500$~GeV is found in
the ZEUS analysis to be 
\begin{equation*}
\alpha_s(M_Z) = 0.1212 \pm0.0017({\rm stat.})^{+0.0023}_{-0.0031}
({\rm exp.})^{+0.0028}_{-0.0027} ({\rm theor.}). 
\label{as:subjet}
\end{equation*}
The experimental error (exp.) is dominated by the uncertainty on the energy
scale for the jet measurement. The largest contribution to the
theoretical uncertainty (theor.) is given by a residual dependency on
the renormalization scale $\mu_r$ which corresponds to uncertainties
due to the contributions of terms beyond next-to-leading order.

\begin{figure}[t]
\center
\epsfig{file=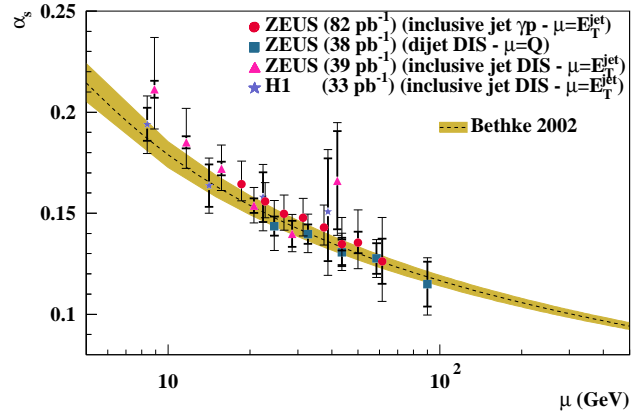,width=0.8\textwidth}
\caption{
Various results on $\alpha_s$ mesurements in DIS as a function of a
mass scale $\mu$ compared with results from global fits.
}
\label{fig:asrun_dis} 
\end{figure}

In Figure~\ref{fig:asrun_dis} the result is displayed as a function of a mass scale $\mu$
together with other values of $\alpha_s$ extracted from DIS-jet
data~\cite{oas}. 

The expected running of $\alpha_s$ as a function of $\mu$
is clearly visible. In addition the figure demonstrates the
compatibility of the resulting $\alpha_s$ values with those obtained
in global fits~\cite{bethke}.

\section{Jet Substructure and Determination of $\alpha_s$}

Jets appear as a collimated spray of particles which are combined to by
dedicated algorithms. These particles are the end point of a cascade
of succsssive particle emissions from the hard interaction to the
hadronic final state. 
\begin{wrapfigure}[22]{l}{5.0cm}
\epsfig{file=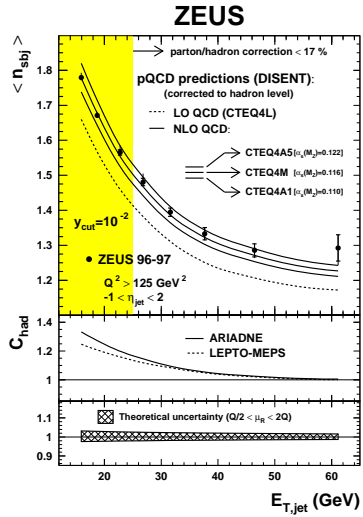,width=0.5\textwidth}
\caption{Subjet multiplicites for different transverse energies
of jets. Jets were identified in the labaratory frame.  
}
\label{fig:sj} 
\end{wrapfigure}
The development of the cascade is governed
by the strong coupling constant $\alpha_s$. An attempt is made to
resolve subjets using dedicated algorithms within the identified 
jets~\cite{zeus:sj}. 
The amount of subjets is expected to depend 
on $\alpha_s$. Fig.~\ref{fig:sj} shows the number of subjets as a function 
of the jet-$E_T$ measured in the laboratory frame. The number of 
subjets decreases as the transverse energy of the jet increases. The
transverse energy of the jet sets the scale for $\alpha_s$. Thus, the
probablity to radiate partons is small at large transverse
energies. The data are well described by NLO-QCD calculations
employing different parton pdfs featuring slightly different values
of $\alpha_s(M_Z)$. For $E_T$ larger than 30~GeV the hadronization
corrections become small allowing for a QCD analysis to determine 
$\alpha_s$ from subjet multiplicites. The resulting value is found by
ZEUS to be
\begin{equation*}
\alpha_s(M_Z) = 0.1187 \pm0.0017({\rm stat.})^{+0.0024}_{-0.0009}
({\rm exp.})^{+0.0093}_{-0.0076} ({\rm theor.}). 
\label{as:sjet}
\end{equation*}

\section{3-jet Cross Sections}
\begin{figure}[t]
\center
\epsfig{file=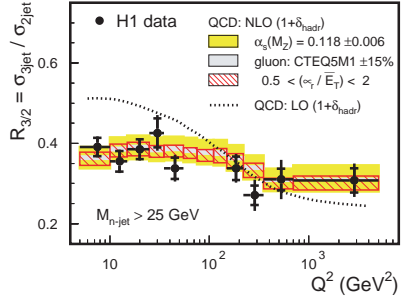,width=0.46\textwidth}
\caption{The observable $R_{3/2}$ as a function of $Q^2$ compared
with NLO-QCD predictions.
}
\label{fig:r32} 
\end{figure}

3-jet cross sections are well suited for an extraction
of $\alpha_s$ because the lowest order contribution to this event
class is proportional to $\alpha^2_s$. The sensitivity to
uncertainties due to proton pdf's can be reduced by building the ratio
$R_{3/2}$, {\it i.e.} the ratio of 3-jet to dijet cross sections. A
measurement of this observable is shown in Fig.~\ref{fig:r32}. 
While a minor
sensitivity to variations of the pdf's is observed the ratio is
very sensitive to small variations of $\alpha_s$ which underlines
the potential of this observable for future determinations of $\alpha_s$.

\section{Conclusion and Outlook}
\begin{wrapfigure}[11]{l}{5.5cm}
\epsfig{file=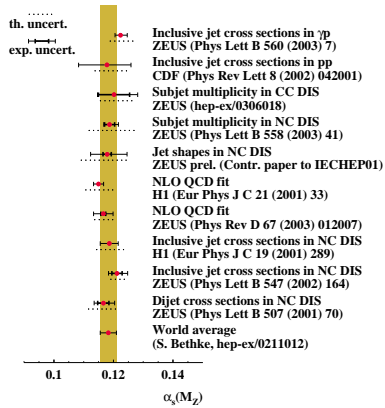, width=0.52\textwidth}
\caption{$\alpha_s(M_Z)$ values obtained in DIS together with results
from $p\bar{p}$-collisions and the world average.}
\label{fig:asmz}
\end{wrapfigure}
The analysis of jet events in DIS allows for precise measurements of
the strong coupling constant $\alpha_s$. A compilation of results is 
given in Fig.~\ref{fig:asmz}. They have a major impact on the current
world average value of $\alpha_s$. Ongoing analysis of HERA
I~\cite{3jet-zeus} data as well
as new data expected from HERA II open the possibility for $\alpha_s$
determinations including 3-jet cross sections.    
\vspace{2.2cm}

\end{document}